\documentclass[aps,prx,floatfix,twocolumn,longbibliography,nofootinbib]{revtex4-2}

\usepackage[utf8]{inputenc}
\usepackage[T1]{fontenc}
\usepackage{physics}
\usepackage{graphicx}
\usepackage{amsmath, amsfonts, amssymb}
\usepackage{bbm, bm}

\usepackage[breaklinks=true, colorlinks]{hyperref}
\hypersetup{linkcolor=red, citecolor=blue}

\usepackage[dvipsnames]{xcolor}
\usepackage{makecell}
\usepackage{upgreek}
\usepackage{nicefrac}

\begin{document}

\title{Bounding speedup of quantum-enhanced Markov chain Monte Carlo}

\author{Alev Orfi}
\affiliation{Center for Computational Quantum Physics, Flatiron Institute, 162 5th Avenue, New York, NY 10010, USA}
\affiliation{Center for Quantum Phenomena, Department of Physics, New York University, 726 Broadway, New York, New York 10003, USA}

\author{Dries Sels}
\affiliation{Center for Computational Quantum Physics, Flatiron Institute, 162 5th Avenue, New York, NY 10010, USA}
\affiliation{Center for Quantum Phenomena, Department of Physics, New York University, 726 Broadway, New York, New York 10003, USA}

\date{\today}

\begin{abstract}

Sampling tasks are a natural class of problems for quantum computers due to the probabilistic nature of the Born rule. Sampling from useful distributions on noisy quantum hardware remains a challenging problem. A recent paper [Layden, D. et al. Nature 619, 282–287 (2023)] proposed a quantum-enhanced Markov chain Monte Carlo algorithm where moves are generated by a quantum device and accepted or rejected by a classical algorithm. While this procedure is robust to noise and control imperfections, its potential for quantum advantage is unclear. Here we show that there is no speedup over classical sampling on a worst-case unstructured sampling problem. We present an upper bound to the Markov gap that rules out a speedup for any unital quantum proposal. 
\end{abstract}

\maketitle

\emph{Introduction.}
Quantum sampling algorithms offer a promising application for small-scale quantum devices. Sampling from certain distributions, achievable with limited qubits, is provably challenging for classical methods \cite{hangleiter2023RandomCircuitSampling}. These complexity results have motivated impressive experimental demonstrations on current noisy processors \cite{arute2019google,zhong2020quantum,wu2021strong,zhong2021phase,zhu2022quantum,madsen2022quantum}. The distributions considered often do not possess immediate practical relevance. Quantum algorithms for sampling from a Boltzmann distribution, a much more applicable task, have been shown to scale favourably as compared to their classical counterparts for certain problems \cite{Szegedy,somma2008quantum,wocjan2008speedup,poulin2009sampling,bilgin2010preparing, temme2011quantum,yung2012quantum,montanaro2015quantum,chowdhury2016quantum,harrow2020adaptive, lemieux2020efficient,arunachalam2022simpler,rall2023thermal,chen2023efficient}. However, these algorithms rely on large, fault-tolerant quantum computers. There has been a recent effort to investigate sampling from a Boltzmann distribution within the limitations of current near-term devices \cite{wild2021quantum,wild2021long,layden2023quantum,zhang2023dissipative,ding2308single}. 

Let $S$ be the configuration space of a system of $N$ discrete spin variables $x_i = \pm 1$. Here we are focused on the task of sampling from the Boltzmann distribution of a classical Hamiltonian $H_c$, of these spin variables. The probability that the system will be in the configuration $x$ at a given temperature $T=1/\beta$ is given by the Boltzmann distribution, 
\begin{equation}\label{eq:boltzmann}
    \pi(x) = \frac{1}{\mathcal{Z}}e^{-\beta H_c(x)}.
\end{equation}
Here $\mathcal{Z}$ is the partition function of the system, 
\begin{equation}\label{eq:partition}
    \mathcal{Z} = \sum_{\{x\}} e^{-\beta H_c(x)},
\end{equation} 
a typically intractable quantity. Sampling from a Boltzmann distribution is ubiquitous in statistical physics, as it is often only possible to estimate thermodynamic quantities using generated samples from this distribution. However, this sampling task has broader significance in fields such as machine learning and optimization. 

A recently proposed near-term quantum algorithm, quantum-enhanced Markov chain Monte Carlo, maintains convergence to a target equilibrium distribution, even with errors in the quantum evolution, making it well-suited for implementation on a near-term quantum device~\cite{layden2023quantum}. Although this algorithm showcases promising error resilience, there is no formal proof of advantage over classical methods. Numerical results presented in Ref.~\cite{layden2023quantum} indicate empirically a polynomial speedup. However, the numerical studies are limited to extremely small systems of less than ten spins for a problem that is known to have large finite-size effects. Without an understanding of the possible mechanism underlying the speedup, it is unclear whether this observed improvement over the classical method would persist at larger scales or for different systems. Here we present a simple example where the quantum-enhanced Markov chain Monte Carlo algorithm has provably no advantage over classical sampling. 

\emph{Classical Sampling.}
Markov chain Monte Carlo (MCMC) is the most common method for sampling from a desired distribution, bypassing the need to explicitly compute $\pi(x)$. MCMC creates a Markov chain, a process where the next configuration is chosen with a fixed probability dependent only on the current configuration of the chain. A Markov chain is specified with a stochastic transition matrix $P$ whose elements describe the probability of moving between any two configurations. Sufficient conditions for convergence to a stationary distribution are that the chain is irreducible, aperiodic, and that it satisfies the detailed balance condition \cite{levin_MarkovChainsMixingTime}. 

Often convergent chains are constructed through a composition of two steps, a proposal step and an acceptance step. If the chain's state is some configuration $y$, a new configuration $x$ is proposed with probability $Q(x|y)$. This new configuration is accepted as the new state of the chain, with probability $A(x|y)$, known as the acceptance probability. As a result, off-diagonal elements of the transition matrix $P$ are the product of these two probabilities,
\begin{equation}
    P(x,y) = Q(x|y)A(x|y).
\end{equation}
The performance of an MCMC algorithm is determined by the rate of convergence to the steady state distribution, defined as the number of steps required for the total variation distance to the stationary state $\pi$ to be $\epsilon$-small. The stationary distribution $\pi$ is associated with the eigenvalue 1 of $P$. Since the distribution of the chain at time $t$ is simply the repeated application of the transition matrix $P$, the spectral gap $\delta = 1-|\lambda_2|$ determines the rate of convergence to the stationary state. More precisely, this property can be used to bound the mixing time $t_{\text{mix}}$,
\begin{equation}\label{eq:mixingTimeBound}
   (\delta^{-1}-1)\ln\left(\frac{1}{2\epsilon}\right)\leq t_{\text{mix}} \leq \delta^{-1}\ln\left(\frac{1}{\epsilon \pi_{\min}}\right)
\end{equation}
where $\pi_{\min} = \min_{x\in S}\pi(x)$ \cite{levin_MarkovChainsMixingTime}.

%The distance between a Markov chain at step $t$ and the distribution $\pi$ is,
%\begin{equation}\label{eq:mixing_time}
%    d(t) = \max_{x\in S}||P^t(x,\cdot)-\pi||_{TV}.
%\end{equation}
%where $||\cdot||_{TV}$ denotes the total variation distance. 

%The mixing time of a Markov chain is defined as the number of steps required for the distance to the stationary state $\pi$ to be $\epsilon$-small \cite{levin_MarkovChainsMixingTime},
%\begin{equation}\label{eq:mixing_time}
%    t_{\text{mix}}(\epsilon)  = \inf\{t\geq 0: d(t)\leq \epsilon\}.
%\end{equation}

\emph{Quantum-Enhanced MCMC.}
The quantum-enhanced Markov chain Monte Carlo algorithm is a Markov chain on the classical configuration space where new configurations are proposed through a measured quantum evolution. Specifically, the current classical state of the Markov chain is prepared as a computational basis state $\ket{x}$, evolved unitarily, and then measured in the computational basis. This procedure gives the following proposal probability, 
\begin{equation*}
    Q(x|y) = |\langle x|U|y\rangle|^2.
\end{equation*}
The focus will be on evolution of the form $U=e^{-iHt}$ with the following time-independent Hamiltonian, 
\begin{equation}\label{eq:ham}
    H = H_c + hH_{\text{mix}}.
\end{equation}
Here, $H_c$ is the classical Hamiltonian defining the desired Boltzmann distribution, $H_{\text{mix}}$ the quantum mixing term, and $h$ is a tunable parameter.
Once a new configuration is found, it is then accepted or rejected classically with Metropolis-Hastings acceptance probability, 
\begin{equation}\label{eq:HMacceptance}
    A(x|y) = \min\left(1,e^{-\beta(H_c(x)-H_c(y))}\frac{Q(y|x)}{Q(x|y)}\right).
\end{equation}
For any symmetric $H$ (in the computational basis), the proposal probability is symmetric, and Eq.~\eqref{eq:HMacceptance} no longer depends on the proposal strategy and can therefore be easily calculated. This property also ensures that the chain satisfies the detailed balance condition. Additionally, if $Q(x|y)>0$ $ \forall x,y$ the chain is also aperiodic and irreducible, thus it converges to the steady state $\pi$ \cite{layden2023quantum}. Only errors in the evolution that break the needed symmetry of the proposal probability bias the MCMC sampling. These can be mitigated through methods such as state preparation and measurement twirling \cite{layden2023quantum}. Other errors may lead to longer mixing times, but the algorithm will still converge to the wanted distribution. 

The quantum proposal method has two free parameters: $h$, which controls the relative weights of the two Hamiltonian terms, and the evolution time $t$. The algorithm was originally formulated with a proposal in which $h$ and $t$ are chosen randomly in each MCMC step, eliminating the need to optimize the free parameters \cite{layden2023quantum}. The performance of this strategy was explored through exact spectral gap calculation and compared to local and uniform classical proposal strategies for the Sherrington-Kirkpatrick model. In the low-temperature regime, the quantum proposal showed favourable mixing time scaling compared to the classical methods for the numerically accessible system sizes \cite{layden2023quantum}. It is suggested that this speedup results from the quantum evolution proposing states that are close in energy to the previous state of the chain and hence are likely accepted while being decorrelated due to thermalization. It should be noted that the energy variance is $hN$ since we start from a computational basis state, suggesting the field might need to be very weak to keep a small change in energy. In addition, for the system to dephase over states within an energy window of the initial state's energy, one would have to evolve for a time that's inversely proportional to the width of the energy window. In hard problems, where nearly degenerate states are prevalent, this energy window would be narrow, implying one may need to evolve coherently on the quantum device for an exponentially long time. To probe the behaviour of this algorithm, consider the following simple setting.

\emph{Marked Item Sampling.} Consider the problem of sampling from a classical Hamiltonian with one marked state, 
\begin{equation}\label{eq:markedHam}
    H_c = -\alpha N\ketbra{k}.
\end{equation}
The scaling by the system size is chosen to ensure the Hamiltonian is extensive. The competition between the extensive energy gain by overlapping with the marked state and the extensive entropy from the degenerate manifold of all other states results in a first-order phase transition in the associated Gibbs distribution Eq.~\eqref{eq:boltzmann} at $T_c=\alpha/\log(2)$. At temperature $T >T_c$, the Gibbs measure on the marked item is exponentially small, while at $T<T_c$, the probability of finding the system in the marked state tends to $1$. Consequently, any classical algorithm should take at least $O(2^N)$ time to sample from the Gibbs state below $T_c$ because it can be used to solve the associated unstructured optimization problem. 

The exact spectral gap of certain MCMC methods can be found for this distribution through methods described in Appendix~\ref{appendix:gap_derivation}. For example, the classical MCMC method, where new configurations are chosen at random and accepted with Metropolis-Hastings acceptance probability, has the following spectral gap,
\begin{equation}\label{eq:uniform_gap}
    \delta = \frac{1}{2^N}\sqrt{e^{-2N\beta\alpha}(2^N-1)^2+2e^{-N\beta\alpha}(2^N-1)+1}.
\end{equation}
We will be focused on the low-temperature regime as the quantum-enhanced Markov chain Monte Carlo showed an improvement in mixing time only in that limit. As expected, for large $\beta$, Eq.~\eqref{eq:uniform_gap} scales inversely with the dimension of the state space. 

\emph{Grover Mixing.}
Consider the mixing Hamiltonian of Eq.~\eqref{eq:ham} to be, 
\begin{equation}
    H_{\text{mix}} = N \ketbra{s}, \, {\rm with}\, \ket{s}=\frac{1}{\sqrt{2^N}}\sum_{x=1}^{2^N} \ket{x}
    \label{eq:GroverMix},
\end{equation}
which maps any input classical state to an equal superposition over all states. Again, the factor $N$ is added to make the Hamiltonian extensive. As the marked state is unknown, this choice is well justified as it unbiasedly mixes the classical configurations. In addition, $H_{\rm mix}$ is the generator of Grover's diffusion operator. The spectral gap of the transition matrix associated with this proposal and Metropolis-Hastings acceptance probability can be found exactly, 
\begin{equation}\label{eq:GroverGap}
    \delta =  \left( \frac{h\sin (N\omega t)}{2^N\omega} \right)^2 \big(1+e^{-N\beta\alpha}(2^N-1)\big),
\end{equation}
with $\omega = \frac{1}{2}\sqrt{(\alpha+h)^2-\alpha h2^{2-N}}$.
The derivation of this spectral gap can be found in  Appendix~\ref{appendix:gap_derivation}. In the low temperature limit, $\delta$ scales with $O(2^{-2N})$ for any $t \ll 1/\omega$. For $t\approx \pi/(2 N\omega)$, one can increase the gap by making the frequency as small as possible. At resonance,
\begin{equation}
    h = -\frac{1}{1-2^{-N}}\alpha,
\end{equation}
the frequency becomes exponentially small $\omega= O(1/\sqrt{2^N})$ such that the gap $\delta$ scales as $O(2^{-N})$. Therefore, for all values of $t$ and $h$, this quantum algorithm has no mixing time scaling improvement over the naive classical strategy. 

\begin{figure}
    \centering
    \includegraphics[width=0.95\columnwidth]{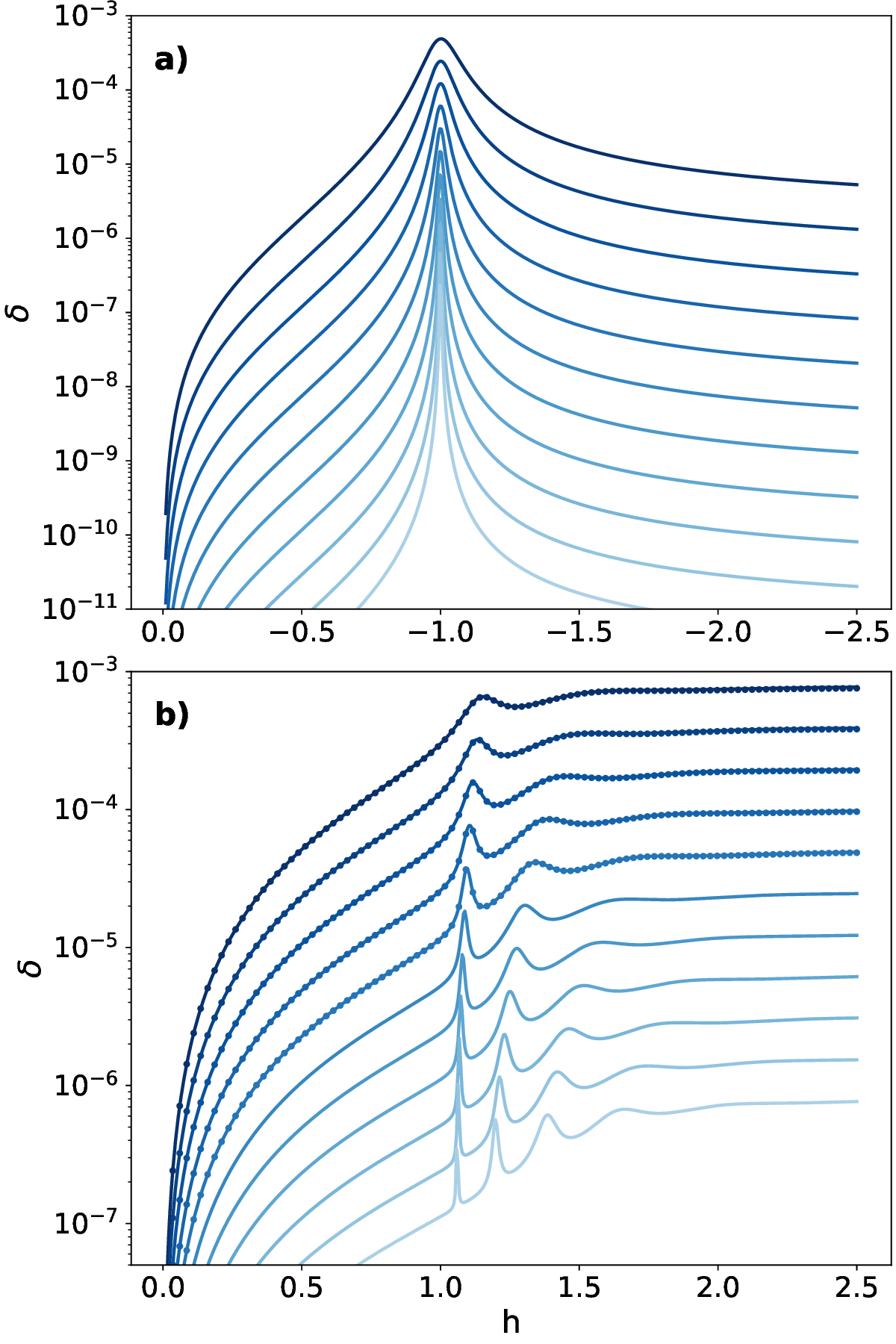}
    \caption{\textbf{a)} Exact time-averaged spectral gap for $N=10-20$ of quantum-enhanced MCMC with a Grover mixing term, i.e. $H_{\text{mix}}$ is given by Eq.~\eqref{eq:GroverMix}. \textbf{b)} Exact spectral gap and bottleneck upper bound of quantum-enhanced MCMC with transverse field mixing term, i.e. $H_{\text{mix}}$ is given by Eq.~\eqref{eq:Xmix}. The exact spectral gap is displayed for $N=10-14$ (dots), while the solid lines show the upper bound of Eq.~\eqref{eq:upperbound} for $N=10-20$. The inverse temperature $\beta=5$ in these figures.}
    \label{fig:gaps-and-bound}
\end{figure}

\emph{General Mixing.}
The Hamiltonian mixing term considered above is non-local and could be difficult to implement experimentally. Moreover, one might be worried that the negative result is a consequence of the particular form of this Grover mixing. Alternatively to the spectral gap, mixing properties can be analyzed by considering the bottlenecks of the chain. These are geometric features that restrict the flow of the chain and ultimately determine the mixing time. The equilibrium flow between two configurations $x$ and $y$ is defined as $E(x,y)=\pi(x)P(x,y)$. For some $S_1,S_2 \subset S$, the flow from $S_1$ to $S_2$ is 
\begin{equation}
    E(S_1,S_2) = \sum_{x\in S_1, y\in S_2}E(x,y).
\end{equation}
After a minimization over possible sets $S_1$, the flow from $S_1$ to its complement, relative to the product of the equilibrium size of the sets, upper bounds the spectral gap \cite{levin_MarkovChainsMixingTime,berestycki_MixingTimes},
\begin{equation}\label{eq:upperbound}
    \delta \leq \min_{S_1: \pi(S_1)\leq 1/2} \frac{E(S_1,S_1^c)}{\pi(S_1)\pi(S_1^c)}.
\end{equation}
This bound is an intermediate step in the derivation of Cheeger's inequality which is often utilized to prove the slow mixing of chains \cite{levin_MarkovChainsMixingTime}.

Consider the set $S_1$ to be all configuration states except the marked state $\ket{k}$. This results in the following upper bound on the gap $\delta$,
\begin{equation}
    \delta \leq \frac{1}{2^N-1}\big(1+e^{-N\beta\alpha}(2^N-1)\big)\sum_{x\neq k} Q(k|x)A(k|x).
    \label{eq:cheegerfull}
\end{equation}
This choice of $S_1$ leads to a tight bound on the spectral gap. For the Grover mixing case, the spectral gap Eq.~\eqref{eq:GroverGap} saturates this bound. Additionally, consider a transverse field mixing term,
\begin{equation}
    H_{\text{mix}} = \sum_i\sigma_i^x,
    \label{eq:Xmix}
\end{equation}
which is the Hamiltonian term originally analyzed when this algorithm was proposed \cite{layden2023quantum}. As shown numerically in Fig.~\ref{fig:gaps-and-bound}, this upper bound is tight to the exact spectral gap and allows for exploration of the behaviour of the chain for large system sizes.   

This upper bound makes the limitations of this quantum proposal strategy evident. For any proposal probability generated through a measured unitary evolution, $Q(k|x)$ is symmetric and the acceptance probability $A(k|x)=1$, hence
\begin{equation}
    \sum_{x\neq k } Q(k|x)A(k|x) \leq 1.
    \label{eq:sumQ}
\end{equation}
This condition drastically limits the effectiveness of this algorithm. Specifically, in the low-temperature regime, the spectral gap scales as $O(2^{-N})$ for any unitary proposal. This analysis can also be extended to the more general setting where new configurations are proposed through an unital quantum channel, which are affine combinations of unitary channels~\cite{mendl09}, 
\begin{equation}
    C(\rho)=\sum_i \lambda_i U_i \,\rho\, U_i^\dagger.
\end{equation}
Consequently the proposal probability $Q(x|y)=\sum_i \lambda_i Q_i(x|y)$. Since the acceptance probability $0\leq A\leq 1$, one can upper bound Eq.~\eqref{eq:cheegerfull} by setting $A=1$, and the double stochasticity of unital channels thus implies Eq.~\eqref{eq:sumQ}. As such, there is no scaling improvement over the uniform classical MCMC method for any unital channel. 

\emph{Conclusion.} In this letter, we analyzed the mixing properties of a quantum-enhanced Markov chain by focusing on a worst-case problem of sampling from a Gibbs distribution of an unknown rank-1 Hamiltonian. At low temperatures, the Gibbs measure is mostly supported on the unknown target state and a Markov chain that samples from the Gibbs measure can thus be used to solve the associated unstructured optimization problem. The latter implies any classical sampling algorithm takes $\Omega(2^N)$ time. Running a quantum computer to generate proposals through an unital channel does not improve the scaling. We have shown this by explicit calculation and by an analysis of the bottleneck of the chain. Future work may extend our bottleneck analysis to problems that are more structured, for which speedup is not ruled out. 

\emph{Acknowledgments}
The authors are grateful for ongoing support through the Flatiron Institute, a division of the Simons Foundation. D.S. is partially supported by AFOSR (grant no. FA9550-21-1-0236) and NSF (grant no. OAC-2118310).

\bibliographystyle{naturemag}
\bibliography{references}

\onecolumngrid
\appendix
\section{Spectral Gap Derivation}
\label{appendix:gap_derivation}
Let $\ket{X_\perp}$ denote an equal superposition state over all but the marked item $\ket{k}$, 
\begin{equation}
    \ket{X_\perp} = \sum_{n\neq k}\frac{1}{\sqrt{2^N-1}}\ket{n}.
\end{equation}
The Hamiltonian under which the system will evolve can be written in terms of spin operators the basis of $\{\ket{X_\perp},\ket{k}\}$,
\begin{equation}
    \begin{split}
         H &= hN\left(1-\frac{1}{2^N}\right)\ketbra{X_\perp}+N\left(-\alpha + \frac{h}{2^N}\right)\ketbra{0} + hN\frac{\sqrt{2^N-1}}{2^N}\Big(\ketbra{X_\perp}{0}+\ketbra{0}{X_\perp}\Big),\\
        &= \frac{N}{2}(h-\alpha)I + \gamma \hat{n}\cdot\Vec{\sigma},
    \end{split}
\end{equation}
where the vector $\hat{n}$ is defined as
\begin{equation}
    \hat{n} = \frac{N}{\gamma}\left(\frac{1}{2}(h+\alpha)-\frac{h}{2^N},h\frac{\sqrt{2^N-1}}{2^N},0\right)
\end{equation}
with 
\begin{equation}
    \gamma = \frac{N}{2}\sqrt{(\alpha+h)^2-\alpha h2^{2-N}}.
\end{equation}
In this form, the time evolution of the system can easily be found,
\begin{equation}
    U = I - \ketbra{0} - \ketbra{X_\perp}+ e^{-it\varphi}\left(\cos(\gamma t)I-i\sin(\gamma t)\Vec{n}\cdot \Vec{\sigma}\right),
\end{equation}
with $\varphi = N(h-\alpha)/2$. In order to construct the transition matrix $P$, the proposal probabilities between classical states are needed. Consider the case where $x \neq k$,
\begin{equation}
    \begin{split}
        Q(k|x) &= |\bra{x}U\ket{k}|^2,\\
    &= \frac{1}{\gamma^22^{2N}}N^2h^2\sin^2(\gamma t).
    \end{split}
\end{equation}
The probability to propose an unmarked state $x$ while in a different unmarked state $y$ is,
\begin{equation}
    \begin{split}
        Q(x|y) &= |\bra{y}U\ket{x}|^2\\
    &= \frac{1}{(2^N-1)^2}\Big[1-2\cos(\varphi t)\cos(\gamma t)+\cos^2(\gamma t)+2n_z \sin(\varphi t)\sin(\gamma t)+n_z^2\sin^2(\gamma t)\Big]\\
    &= \frac{K}{(2^N-1)^2}.
    \end{split}
\end{equation}
Elements of the transition matrix describe the probability of moving from one state $y$ to another $x$. Specifically,
\begin{equation}\label{eq:Pconstruct}
    P(y,x) = \begin{cases}
     Q(x|y)A(x|y), & \text{if}\ x\neq y\\
     1-\sum_{\{z \neq y\}} Q(x|z)A(x|z), & \text{if}\ x=y
    \end{cases},
\end{equation}
where $A(x|y)$ denotes the Metropolis-Hastings acceptance probability. Due to the structure of this distribution, this matrix only has five unique elements. First, the probability of moving from an arbitrary state to the marked state is, 
\begin{equation}
    P(x,k) = Q(k|x),
\end{equation}
whereas moving from the marked state to an arbitrary state is
\begin{equation}
    P(k,x) = Q(x|k)e^{-N\beta\alpha} = Q(k|x)e^{-N\beta\alpha}.
\end{equation}
The probability of staying in the marked state is then fixed through the stochasticity requirement of $P$, 
\begin{equation}
    P(k,k) = 1-(2^N-1)Q(k|x)e^{-N\beta\alpha}.
\end{equation}
Moving between unmarked states is not penalized by the Metropolis-Hastings probability, 
\begin{equation}
    P(y,x) = Q(x|y),
\end{equation}
and again requiring the matrix to be stochastic gives, 
\begin{equation}
    P(x,x) = 1 - (2^N-1)Q(x|y) - Q(k|x).
\end{equation}
The transition matrix, ordered such that the marked item is the first index, is then, 
\begin{equation}
    P= \begin{bmatrix}
        P(k,k)& P(k,x)  &P(k,x) & ...\\
        P(k,x) & P(x,x) & P(y,x) &...\\
        P(k,x) & P(y,x) & P(x,x) &...\\
        \vdots
    \end{bmatrix}\\ = \Tilde{P} +(P(x,x)-P(y,x))I
\end{equation}
Where $\Tilde{P}$ has the same spectral gap as $P$. Additionally, a similarity transformation on $\Tilde{P}$ gives,
\begin{equation}
    \begin{bmatrix}
        P(k,k)+P(y,x)-P(x,x)  &P(k,x)e^{N\beta\alpha/2}&P(k,x)e^{N\beta\alpha/2}& ...\\
        P(k,x)e^{N\beta\alpha/2}& P(y,x) & P(y,x) &...\\
        P(k,x)e^{N\beta\alpha/2}& P(y,x) & P(y,x) &...\\
        \vdots
    \end{bmatrix}
    =A\ketbra{0} + B\Big(\ketbra{X_\perp}{0}+\ketbra{0}{X_\perp}\Big) +C\ketbra{X_\perp}.
\end{equation}
This reduced the problem into a two-by-two matrix with elements, 
\begin{equation}
    A=\frac{K}{2^N-1}+\frac{N^2h^2\sin^2(\gamma t)(2^N-1)}{2^{2N}\gamma^2}(1-e^{-N\beta\alpha}),
\end{equation}
\begin{equation}
    B=e^{-N\beta\alpha}\frac{N^2h^2\sin^2(\gamma t)\sqrt{2^N-1}}{2^{2N}\gamma^2},
\end{equation}
\begin{equation}
    C = \frac{K}{2^N-1}.
\end{equation}
The spectral gap of $P$ is then, 
\begin{equation}
    \delta = \sqrt{4B^2+C^2-2CA+A^2} = \left( \frac{h\sin (N\omega t)}{2^N\omega} \right)^2 \big(1+e^{-N\beta\alpha}(2^N-1)\big)
\end{equation}
with $\omega = \gamma/N = \frac{1}{2}\sqrt{(\alpha+h)^2-\alpha h2^{2-N}}$. 

\end{document}